\documentclass[useAMS,usenatbib]{mn2e}
\usepackage{times}
\usepackage{graphicx}
\usepackage{pdflscape}
\usepackage{epsfig}
\usepackage{ctable}
\usepackage{natbib}
\usepackage{romannum}
\usepackage{array}
\newcolumntype{+}{>{\global\let\currentrowstyle\relax}}
\newcolumntype{^}{>{\currentrowstyle}}
\newcommand{\rowstyle}[1]{\gdef\currentrowstyle{#1}%
  #1\ignorespaces
}

\newcommand{\rxte}{\textit{RXTE}}

\newcommand{\sw}{\textit{Swift}}

\def\cygx1{Cygnus~X$-$1}
\def\gx{GX~339$-$4}

\def\msun{$M_{\odot}$}

\def\ergcms{erg~cm$^{-2}$~s$^{-1}$}
\def\kms{km~s$^{-1}$}
\def\wm2{W~m$^{-2}$}

\def\cm2{cm$^{-2}$}
\def\se1{s$^{-1}$}

\def\Ave{A_{\rm V}}

\usepackage{multirow}
\usepackage{array}
\usepackage{appendix}
\usepackage{booktabs}
\usepackage{multirow}

\title[Optical and near-infrared spectroscopy of \gx]{Optical and
  near-infrared spectroscopy of the black hole \gx\\II. The
  spectrocopic
  content in the low/hard and high/soft states \thanks{Based on observations
  performed with European Southern Observatory (ESO) Telescopes at the
Paranal Observatory under programmes ID 284.D-5056 and 285.D-5007}}
\author[F. Rahoui et al.]{Farid Rahoui$^{1,2}$\thanks{E-mail:
frahoui@eso.org}, Mickael Coriat$^{3}$ and Julia C. Lee$^{2}$\\
$^{1}$European Southern Observatory, K. Schwarzschild-Str. 2, 85748
Garching bei M\"unchen, Germany\\
$^{2}$Department of Astronomy, Harvard University, 60 Garden street,
Cambridge, MA 02138, USA\\ 
$^{3}$Astronomy, Astrophysics, Cosmology and Gravity Centre, 
University of Cape Town, Private Bag X3 Rondebosch, 7701 South Africa
}
\begin{document}
\date{Accepted 2014 May 14.  Received 2014 May 13; in original form 2014 January 20}
\pagerange{\pageref{firstpage}--\pageref{lastpage}} \pubyear{2014}
\maketitle
\label{firstpage}

\begin{abstract}
As a complement to our optical and near-infrared study of the
continuum properties of \gx\ in the two hard and one soft state
observations made by the ESO/VLT FORS2 and ISAAC in early 2010, we
report here on the results of our spectral line analysis for the same
observations. In the soft state, the presence of strong Balmer,
Paschen and Brackett emission lines points to the optical and
near-infrared spectra stemming from the irradiated chromosphere of the
optically thick and geometrically thin accretion disc. Most of these
\ion{H}{1} features are still detected in emission in both hard states
but are veiled by the compact jets continuum. We also confirm
the presence of a broad H{\small $\beta$} absorption feature,
prominent in the soft state and shallower in the first hard state,
which we argue forms in the deep layers of the optically thick
accretion disc. However, this trough is absent in the second hard
state, a likely consequence of the formation
of a geometrically thick extended envelope that arises above the
  disc plane and eventually enshrouds the region where the H{\small
    $\beta$} absorption feature forms. We detect this envelope
through the presence of a broad Pa{\small $\beta$} emission
line, which is constant during the first hard state but correlates
with the underlying continuum during the second hard state, pointing
to changing physical properties. We consider that this behaviour may
be consistent with the launch of a thermally-driven accretion disc wind
during the second hard state.

\end{abstract}


\section{Introduction}

Detailed analysis of the optical and near-infrared (near-IR)
spectroscopic content of microquasars can be challenging, as many of
them are distant and suffer from a high extinction from the
interstellar medium (ISM) along their line-of-sight (LOS). Previous 
studies however revealed that some exhibited very rich
spectra during outbursts that gave strong insights into the physical
processes at play \citep[see e.g.][]{1995Callanan, 1997Bandy, 2000Soria,
  2001Dubus}. Although \gx\ has been extensively studied in the
optical
\citep{1986Makishima,1999Smith,1999Soria,2001Shahbaz,2001Wu,2002Cowley,
  2003Buxton,2003Hynes,2004Hynes}, 
there has been very little near-infrared
(near-IR) spectral investigations \citep[][hereafter Paper~I]{2012Rahoui}. 
Yet, this source is probably one of the best targets for  
investigating the near-IR spectral characteristics of
microquasars during outburst due to the relatively low extinction
along its LOS \citep[$\Ave=3.25\pm0.5$,][]{2011Gandhi}, its
brightness ($J\sim12-15$) and because the emission of the 
companion star is mostly dominated by that of the accretion disc
in the soft state (SS) or the compact jets in the hard state (HS) \citep{2001Shahbaz}. 
\begin{figure*}
\begin{center}
\includegraphics[width=15cm]{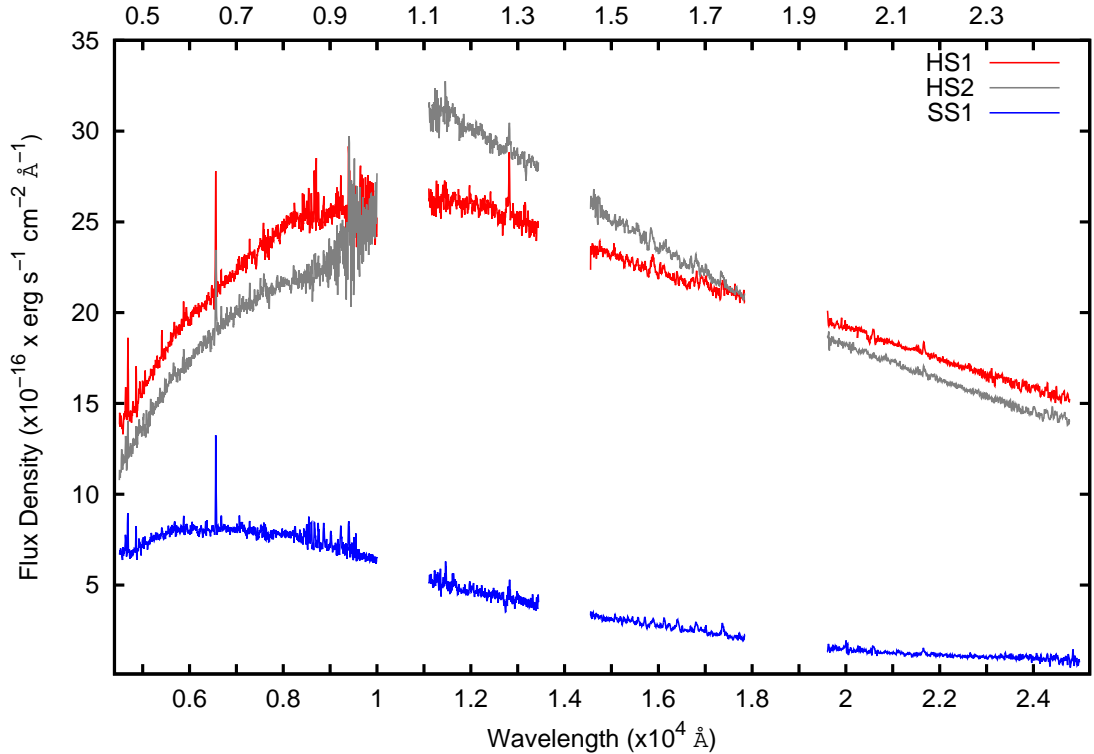}
\caption{\small Flux-calibrated spectra of \gx\ during HS1 (red), HS2
  (grey) and SS1 (blue). The spectra are not corrected for
the ISM extinction along the LOS of the source.}
\label{contspec}
\end{center}
\end{figure*}
In Paper~I, we presented results based on the
analysis of the optical/near-IR (OIR) continuum of \gx, supported by
simultaneous X-ray and radio data. The source was observed twice in
the hard state (MJD~55261.4 and MJD~55262.4, hereafter HS1 and HS2,
  respectively) and once in the soft state (MJD~55307.3,
  hereafter SS1). We showed that the OIR spectrum stemmed from 
\newline
(1) a high level of disc irradiation during 
SS1, with an excess possibly due to the companion star 
at longer wavelengths, 
\newline
(2) a combination of optically thin synchrotron emission
from the compact jet with a steep UV/optical component during HS1,
\newline
(3) more complex emission during HS2; the optical spectrum, fainter
than during HS1, was consistent with disc irradiation, but a strong
near-IR excess was present and we suggested that it might have been
due to more ``exotic'' processes in the jet and/or corona
\citep{2005Markoff, 2009Peer, 2011Veledina}. 
\newline
The near-IR continuum was also 
found to be variable on 20~s timescales
in both HS observations but constant during SS1, suggesting a
jet-related origin to these variations.

In this paper, we report on the second part of our study, which
focuses on the spectroscopic content in the optical and
near-IR domains and aims at (1) making the first detection of
spectral lines in the near-IR and (2) understanding the origin and
properties of the spectral features in both the HS and SS. The paper
is organised in the following way. Section~\ref{secobs} details the
specific reduction of the near-IR spectra we performed to avoid
contamination by the telluric standards. The spectra are presented in
Section~\ref{secres} and analysed in Section~\ref{secan}. We discuss
the results in Section~\ref{secdis} and conclude in
Section~\ref{seccon}.

\section{Observations and data reduction}\label{secobs}

The observations are presented in detail in Paper~I (\textsection~2)
and we refer the reader to that paper for an in-depth description of
the data reduction and flux calibration. Whereas Paper~I focused
primarily on the continuum, here we investigate more specifically 
  on the OIR spectral features. In the near-IR, the
data reduction process introduces a contamination by the spectroscopic
features of the standard star, which must therefore be removed as much
as possible prior to the telluric division. The standard stars which
were observed are all giant or main sequence B3--9, as characterised in
ISAAC low-resolution spectra by the sole presence of the Paschen and
Brackett series in their near-IR spectra \citep[see e.g.][]{1997Wallace,
  1998Meyer, 2000Wallace}. In the \textit{J}- and \textit{K}-bands,
Pa{$\beta$} and Br{$\gamma$} were isolated enough from any atmospheric
absorption dip to be simply removed through interpolation with
Gaussians. Unfortunately, this could not be easily done in the
\textit{H}-band because the presence of many Brackett lines between
15000 and 16000~\AA\ as well as the overlapping of Br13 and Br15 with
broad atmospheric absorption features made the fitting of the continuum
difficult. In the \textit{H}-band the solution then is to remove the
spectroscopic lines of the standard stars by
dividing their observed spectra (including atmospheric contribution)
by the corresponding Kurucz ones normalised to unity
\citep{1979Kurucz}. The Kurucz spectra were first rebinned to ISAAC
resolution and modified by limb darkening and rotational broadening
\citep[see][]{2008Gray} so that their Br10, Br11 and Br12 lines
matched those in the observed spectra. We then checked the
  resulting feature-free standard star spectra to ensure no measurable
  residuals.
\begin{table*}
\begin{center}
\caption{Optical and near-IR lines in \gx\ spectra. ``$\dag$'' marks the features that are difficult to
measure due to their proximity with an atmospheric/ISM absorption dip and/or complex continuum; these
measurements should be taken with caution. ``$\ast$'' means that the
lines are detected but measurements would be unreliable; ``--''
indicates that the lines are not detected at all. Uncertainties are given at $1\sigma$.}
\label{speclines}
\scalebox{0.82}
{\begin{tabular}{+l^c|^c^c^c^c|^c^c^c^c|^c^c^c^c}
\hline\hline
&&\multicolumn{4}{c|}{SS1}&\multicolumn{4}{c|}{HS1}&\multicolumn{4}{c}{HS2}\\
\hline
{Feature} & {$\lambda_{\rm \small L}$}$^a$ & ${F_{\rm cont}}^b$&${\mathring{W}}^c$ &$FWHM^d$ &${F_{\rm line}}^e$ & ${F_{\rm cont}}$&$\mathring{W}$&$FWHM$&${F_{\rm line}}$&${F_{\rm cont}}$&$\mathring{W}$&$FWHM$&${F_{\rm line}}$ \\
\hline
$\dag$H{$\gamma$}&	4342	& $\ast$	&$\ast$		&	$\ast$		&	$\ast$		&	$\ast$	&$\ast$	        &       $\ast$          & $\ast$         &      --	& --                    &  --		        & --\\
\ion{N}{3}	&	4515	& $\ast$	&$\ast$		&	$\ast$ 		&	$\ast$ 		&	1.27	&-0.41$\pm$0.17	&	248$\pm$159	& 0.56$\pm$0.21  &	--	& --			&  --			& --	\\
\ion{He}{2}	&	4542	& $\ast$	&$\ast$		&	$\ast$ 		&	$\ast$ 		&	1.28	&-0.42$\pm$0.20	&	949$\pm$100	& 0.55$\pm$0.22	 &	--	& --			&  --			& --	\\
\ion{N}{3}	&	4640	& 0.69		&-3.12$\pm$0.33	&	1166$\pm$35 	&	2.06$\pm$0.18	&	1.35	&-2.21$\pm$0.24	&	754$\pm$17	& 3.02$\pm$0.12  &      1.06    & -1.21$\pm$0.20	&  708$\pm$18		& 1.29$\pm$0.16 \\
\ion{He}{2}	&	4686	& 0.69		&-4.45$\pm$0.53	&	532$\pm$25 	&	3.02$\pm$0.19	&	1.40	&-4.62$\pm$0.25	&	770$\pm$23	& 6.41$\pm$0.36	 &      1.10	& -2.45$\pm$0.25	&  566$\pm$26		& 2.78$\pm$0.19 \\
?		&	4741	& --		&--		&	-- 		&	--		&	$\ast$	&$\ast$		&	$\ast$		&$\ast$	 	 &	$\ast$	& $\ast$		&  $\ast$		& $\ast$ \\
?		&	4801	& --		&--		&	-- 		&	--		&	$\ast$	&$\ast$		&	$\ast$		&$\ast$	 	 &	$\ast$	& $\ast$		&  $\ast$		& $\ast$ \\
\rowstyle{\bfseries}
H{$\beta+$}	&	4861	& 0.68		&-2.85$\pm$0.70	&	496$\pm$82	&	2.06$\pm$0.40	&	1.47	&-1.64$\pm$0.38	&	206$\pm$86	& 2.37$\pm$0.35	 &      1.22	& -0.79$\pm$0.24	&  256$\pm$84	        & 0.92$\pm$0.27 \\
H{$\beta-$}	&		& 0.70		&2.89$\pm$0.81	&	2400$\pm$255	&	2.00$\pm$0.5	&	1.49	&1.27$\pm$0.32	&	1662$\pm$398	& 1.95$\pm$0.38	 &      1.23	& 0.33$\pm$0.26		&  321$\pm$140		& 0.48$\pm$0.30 \\
\ion{He}{1}	&	4922	& $\ast$	&$\ast$		&	$\ast$		&	$\ast$		&	1.51	&-0.45$\pm$0.12	&	273$\pm$76	& 0.67$\pm$0.19	 &	$\ast$	& $\ast$		&  $\ast$		& $\ast$\\
\ion{O}{2}	&	4942	& $\ast$	&$\ast$		&	$\ast$		&	$\ast$		&	1.51	&-0.54$\pm$0.18	&	918$\pm$94	& 0.87$\pm$0.27	 &	$\ast$	& $\ast$		&  $\ast$		& $\ast$\\
\ion{N}{2}	&	5005	& $\ast$	&$\ast$		&	$\ast$ 		&	$\ast$		&	1.57	&-0.83$\pm$0.35	&	455$\pm$186	& 0.81$\pm$0.25	 &	--	& --			&  --			& --	\\
\ion{He}{1}	&	5017	& $\ast$	&$\ast$ 	&	$\ast$		&	$\ast$		&	$\ast$	&$\ast$ 	&	$\ast$		&	$\ast$	 &	--	& --			&  --			& --	\\
\ion{He}{2}	&	5413	& 0.80		&-0.56$\pm$0.33	&	202$\pm$84 	&	0.48$\pm$0.17	&	$\ast$	&$\ast$ 	&	$\ast$		&	$\ast$	 &	--	& --			&  --			& -- 		\\
\ion{He}{1}	&	5876	& $\ast$	&$\ast$ 	&	$\ast$		&	$\ast$		&	$\ast$	&$\ast$ 	&	$\ast$		&	$\ast$	 &  	$\ast$	& $\ast$		&  $\ast$		& $\ast$ \\
%$[$\ion{O}{1}$]$&	6364	& $\ast$	&$\ast$		&	$\ast$ 		&	$\ast$		&	--	&--		&	--		& --		 &	--	& --			&  --			& -- 		\\
?		&	6475	& $\ast$	&$\ast$		&	$\ast$ 		&	$\ast$		&	$\ast$	&$\ast$		&	$\ast$		&$\ast$	 	 &	--	& --			&  --			& -- 	\\
?		&	6505	& $\ast$	&$\ast$		&	$\ast$ 		&	$\ast$		&	$\ast$	&$\ast$		&	$\ast$		&$\ast$	 	 &	--	& --			&  --			& -- 	\\
\rowstyle{\bfseries}
H{$\alpha$}	&	6563	& 0.82		&-8.12$\pm$0.47	&	421$\pm$44  	&	6.79$\pm$0.25	&	2.11	&-5.56$\pm$0.46	&	550$\pm$30	& 12.01$\pm$0.35 &   	1.96 	& -4.53$\pm$0.31	&  704$\pm$53	        & 8.65$\pm$0.29 \\
\ion{He}{1}	&	6678	& 0.82		&-1.27$\pm$0.23	&	444$\pm$33 	&	0.98$\pm$0.12	&	2.16	&-0.77$\pm$0.16	&	541$\pm$36	& 1.77$\pm$0.16  &	2.01	& -0.69$\pm$0.13	&  707$\pm$34		& 1.33$\pm$0.18 \\
\ion{He}{1}	&	7065	& $\ast$	&$\ast$		&	$\ast$ 		&	$\ast$		&	$\ast$	&$\ast$		&	$\ast$		&$\ast$	 	 &	$\ast$	& $\ast$		&  $\ast$		& $\ast$\\
\ion{He}{2}	&	7280	& $\ast$	&$\ast$		&	$\ast$ 		&	$\ast$		& 	$\ast$	&$\ast$		&	$\ast$		&$\ast$	 	 & 	--	& --			&  --			& -- 		\\
Pa (39-3)	&	8252	& $\ast$	&$\ast$		&	$\ast$ 		&	$\ast$		&	$\ast$	&$\ast$		&	$\ast$		&$\ast$	 	 &	$\ast$	& $\ast$		&  $\ast$		& $\ast$ \\
Pa (22-3)	&	8359	& $\ast$	&$\ast$		&	$\ast$ 		&	$\ast$		&	$\ast$	&$\ast$		&  	$\ast$		&$\ast$ 	 &	$\ast$	& $\ast$		&  $\ast$		& $\ast$ 	\\
Pa (18-3)	&	8438	& $\ast$	&$\ast$		&	$\ast$ 		&	$\ast$		&	$\ast$	&$\ast$		&	$\ast$		&$\ast$	 	 &	$\ast$	& $\ast$		&  $\ast$		& $\ast$ \\
$\dag$Pa (16-3)	&	8502	& 0.72		&-1.30$\pm$0.39	&	479$\pm$47 	&	0.94$\pm$0.22	&	2.33	&-0.50$\pm$0.27	&	481$\pm$75	& 1.13$\pm$0.41	 &	2.18	& -0.30$\pm$0.16	&  519$\pm$119		& 0.71$\pm$0.20 \\
\rowstyle{\bfseries}
Pa (15-3)	&	8545	& 0.71		&-1.82$\pm$0.27	&	636$\pm$50 	&	1.33$\pm$0.19	&	2.31	&-0.65$\pm$0.16	&	515$\pm$80	& 1.75$\pm$0.26	 &	2.18	& -0.48$\pm$0.22	&  541$\pm$82	        & 1.45$\pm$0.37 \\
\rowstyle{\bfseries}
Pa (14-3)	&	8598	& 0.70		&-2.21$\pm$0.29	&	621$\pm$52 	&	1.66$\pm$0.27	&	2.30 	&-1.01$\pm$0.20 &	559$\pm$43	& 2.43$\pm$0.37	 &	2.18	& -0.88$\pm$0.21	&  541$\pm$89	        & 1.73$\pm$0.48 \\
\rowstyle{\bfseries}
Pa (13-3)	&	8665	& 0.70		&-2.27$\pm$0.35	&	662$\pm$82 	&	1.74$\pm$0.23	&	2.32	&-1.19$\pm$0.19	&	632$\pm$56	& 2.42$\pm$0.35  &	2.34	& -0.92$\pm$0.18	&  573$\pm$108	        & 1.70$\pm$0.31 \\
\rowstyle{\bfseries}
Pa (12-3)	&	8750	& 0.69		&-2.30$\pm$0.41	&	609$\pm$43 	&	1.68$\pm$0.24	&	2.35	&-0.90$\pm$0.18	&	601$\pm$93	& 2.12$\pm$0.38	 &	2.34	& -0.64$\pm$0.23	&  594$\pm$99	        & 1.58$\pm$0.36 \\
\rowstyle{\bfseries}
Pa (11-3)	&	8863	& 0.67		&-2.49$\pm$0.38	&	576$\pm$53 	&	1.83$\pm$0.22	&	2.34	&-0.88$\pm$0.21	&	610$\pm$58	& 2.34$\pm$0.29	 &	2.34	& -0.82$\pm$0.31	&  566$\pm$124	        & 1.73$\pm$0.37 \\
\ion{He}{2}	&	8931	& $\ast$	&$\ast$		&	$\ast$ 		&	$\ast$		&	--	&--		&	--		& --		 &	--	& --			&  --			& -- 		\\
$\dag$Pa (10-3)	&	9015	& 0.62		&-2.80$\pm$0.45	&	648$\pm$56 	&	1.77$\pm$0.22	&	2.25	&-1.08$\pm$0.28	&	588$\pm$84	& 2.41$\pm$0.45	 &	$\ast$	&$\ast$	 	 	&$\ast$			& $\ast$	\\
$\dag$Pa (9-3)	&	9229	& 0.60		&-5.82$\pm$1.05	&	703$\pm$87	&	3.67$\pm$0.70	&	2.27	&-2.44$\pm$0.48	&	706$\pm$43	& 5.09$\pm$0.67	 &  	$\ast$	&$\ast$	 	 	&$\ast$			& $\ast$	\\
\rowstyle{\bfseries}
Pa{$\beta$}	&	12818	& 0.41		&-7.00$\pm$0.53 &	490$\pm$84 	&	3.00$\pm$0.25	&	2.51    &-4.91$\pm$0.51	&	891$\pm$61	& 12.91$\pm$0.63 &	2.93	& -2.79$\pm$0.30	&  1105$\pm$43	        & 8.26$\pm$0.42	\\
Br (19-4)	&	15265	& $\ast$	&$\ast$		&	$\ast$ 		&	$\ast$		&	$\ast$	&$\ast$		&	$\ast$		& $\ast$	 &	$\ast$	& $\ast$		&  $\ast$		& $\ast$ \\
$\dag$Br (18-4)	&	15346	& $\ast$	&$\ast$		&	$\ast$ 		&	$\ast$		&	2.30	&-0.30$\pm$0.21	&	687$\pm$116	& 1.03$\pm$0.38	 &	$\ast$	& $\ast$		&  $\ast$		& $\ast$ \\
$\dag$Br (17-4)	&	15443	& 0.28		&-2.42$\pm$0.26	&	343$\pm$80 	&	0.63$\pm$0.11	&	2.28	&-0.44$\pm$0.11	&	482$\pm$111	& 1.32$\pm$0.26	 &	$\ast$	& $\ast$		&  $\ast$		& $\ast$ \\
Br (16-4)	&	15561	& 0.27		&-3.48$\pm$0.25	&	360$\pm$71 	&	0.96$\pm$0.12	&	2.26	&-0.61$\pm$0.19	&	466$\pm$103	& 1.33$\pm$0.20	 &	$\ast$	& $\ast$		&  $\ast$		& $\ast$ \\
$\dag$Br (15-4)	&	15705	& 0.27		&-3.91$\pm$0.32	&	413$\pm$113 	&	1.04$\pm$0.11	&	$\ast$	&$\ast$		&	$\ast$		&$\ast$	 	 &	2.71	& -0.30$\pm$0.11	&  466$\pm$224		& 1.07$\pm$0.33\\
%\rowstyle{\bfseries}
\ion{He}{1}	&	15859	& 0.26		&-1.70$\pm$0.71	&	646$\pm$186 	&	0.47$\pm$0.20	&	2.21	&-0.75$\pm$0.20	&	433$\pm$118	& 2.02$\pm$0.35  &  	2.65	& -0.80$\pm$0.23	&  508$\pm$129		& 2.18$\pm$0.47 \\
\rowstyle{\bfseries}
Br (14-4)	&	15885	& 0.26		&-4.96$\pm$0.55	&	700$\pm$84 	&	1.41$\pm$0.25	&	2.21	&-1.22$\pm$0.37	&	483$\pm$108	& 2.33$\pm$0.47  &  	2.65	& -0.60$\pm$0.19	&  677$\pm$107		& 1.43$\pm$0.32 \\
$\dag$Br (13-4)	&	16114	& 0.25		&-7.00$\pm$0.53	&	553$\pm$95 	&	1.67$\pm$0.23	&	2.17	&-1.35$\pm$0.25	&	534$\pm$86	& 2.65$\pm$0.31	 &	2.59	& -0.39$\pm$0.12	&  478$\pm$88		& 1.02$\pm$0.31 \\
$\dag$\ion{He}{2}&	16246	& 0.25		&-2.28$\pm$0.51 &	492$\pm$88	&	0.54$\pm$0.13	& 	--	&--		&	--		& -- 		 & 	$\ast$	& $\ast$		&  $\ast$		& $\ast$ \\
Br (12-4)	&	16412	& 0.25		&-7.71$\pm$0.67	&	600$\pm$72 	&	2.00$\pm$0.21	&	2.14	&-0.49$\pm$0.11	&	247$\pm$124	& 1.75$\pm$0.27	 &	--	& --			&  --			& -- \\
\rowstyle{\bfseries}
Br (11-4)	&	16811	& 0.23		&-7.86$\pm$0.72 &	621$\pm$83 	&	1.99$\pm$0.19	&	2.10	&-1.14$\pm$0.21	&	476$\pm$85	& 2.73$\pm$0.28  &   	2.41	& -0.62$\pm$0.18	&  527$\pm$87	        & 1.81$\pm$0.42 \\
$\dag$\ion{He}{2}&	16923	& --		&--		&	--		&	--		&	2.09	&-1.12$\pm$0.32	&	426$\pm$122	& 2.30$\pm$0.57	 &	2.38	& -0.31$\pm$0.13	&  544$\pm$122	       	& 0.74$\pm$0.21\\
$\dag$\ion{He}{1}&	17000 	& 0.23		&-4.41$\pm$0.45	&	168$\pm$166 	&	0.96$\pm$0.16	&	2.09	&-0.60$\pm$0.22	&	472$\pm$117	& 1.25$\pm$0.33	 &	2.36	& -0.20$\pm$0.12        &  224$\pm$162	       	& 0.50$\pm$0.19 \\
\rowstyle{\bfseries}
Br (10-4)	&	17367	& 0.22		&-8.20$\pm$0.87&	606$\pm$79 	&	2.51$\pm$0.30	&	2.06	&-1.50$\pm$0.46	&	488$\pm$90	& 3.27$\pm$0.39  &   	2.28	& -0.70$\pm$0.31	&  760$\pm$109	        & 2.53$\pm$0.41 \\
$\dag$\ion{He}{1}&	20586	& 0.14		&-10.01$\pm$1.09&	222$\pm$125 	&	1.36$\pm$0.34	&	$\ast$	&$\ast$		&	$\ast$		& $\ast$	 &	--	& --			&  --			& -- \\	
\ion{He}{1}	&	21126	& 0.13		&-6.85$\pm$1.12	&	574$\pm$161 	&	0.87$\pm$0.23	&	--	&--		&	--		& --		 &	--	& --			&  --			& -- \\	
\rowstyle{\bfseries}
Br{$\gamma$}	&	21661	& 0.13		&-13.46$\pm$1.03&	553$\pm$57 	&	1.71$\pm$0.23	&	1.72	&-1.42$\pm$0.23	&	220$\pm$102	& 2.45$\pm$0.25	 &	1.66	& -1.20$\pm$0.20	&  547$\pm$54 	        & 1.79$\pm$0.25 \\
\hline 
\end{tabular}}
\newline
{{$^a$}{Laboratory wavelength {(\AA)}} \\}
{{$^b$}{Continuum flux} {($\times 10^{-15}$erg cm$^{-2}$ s$^{-1}$ \AA$^{-1}$)}\\}
{{$^c$}{Equivalent widths} {(\AA)}\\}
{{$^d$}{Full-width at half-maximum} (\kms)\\}
{{$^e$}{Line flux} {($\times 10^{-15}$erg cm$^{-2}$ s$^{-1}$)}\\}
\end{center}
\end{table*}

\section{Spectral description}\label{secres}

\begin{figure}
\begin{center}
\includegraphics[width=8.5cm]{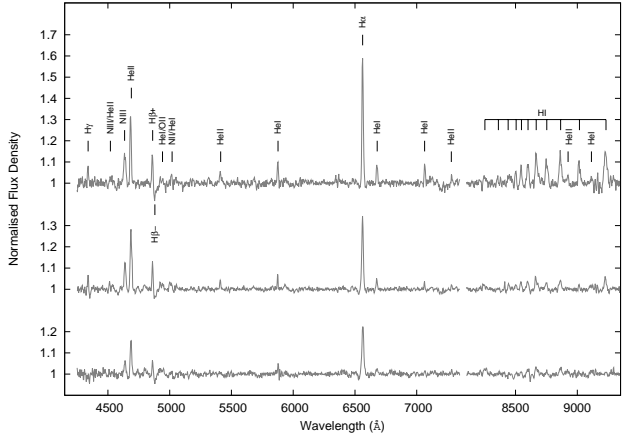}
\caption{\small Continuum-normalised optical spectra of \gx\ for SS1 (top), HS1
  (middle) and HS2 (bottom). All the detected lines are marked, and
  the atmospheric/ISM features were manually removed.}
\label{normvspec}
\end{center}
\end{figure}

\begin{figure}
\begin{center}
\includegraphics[width=8.5cm]{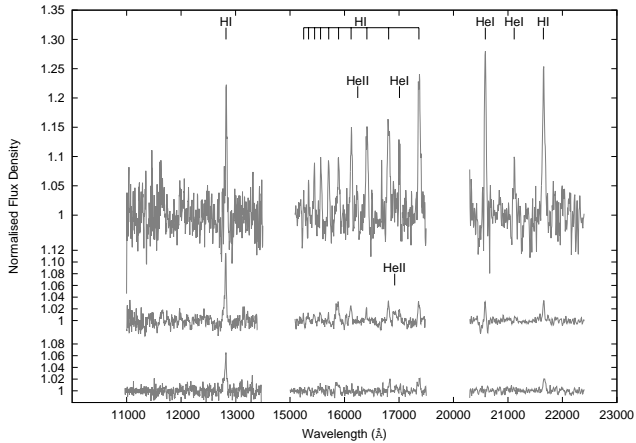}
\caption{\small Continuum-normalised near-IR spectra of \gx\ for SS1 (top), HS1
  (middle) and HS2 (bottom), in each band ({\it J} to {\it K}, left to
  right). All the detected lines are marked.} 
\label{normirspec}
\end{center}
\end{figure}

The resulting OIR flux-calibrated spectra are displayed in
Figure~\ref{contspec}, and the continuum-normalised ones in 
Figures~\ref{normvspec} and \ref{normirspec}. Likewise,
Table~\ref{speclines} lists all the detected 
lines and their characteristics. The full-widths at
half-maximum (FWHMs) were quadratically corrected for the
instrumental broadening which we assessed from the arc spectra, and
the underlying continuum was locally assessed with a first-order
polynomial. The latter being the primary source of inaccuracy, each
measurement was repeated several times with different continuum
placements within the same wavelength range to obtain a set of
values that eventually averaged 
out. The uncertainties listed in Table~\ref{speclines} are therefore
the scatter to the mean rather than the real errors.

\subsection{The optical spectrum}

\begin{figure*}
\begin{center}
\begin{tabular}{ccc}
\includegraphics[height=3.83cm,width=5.5cm]{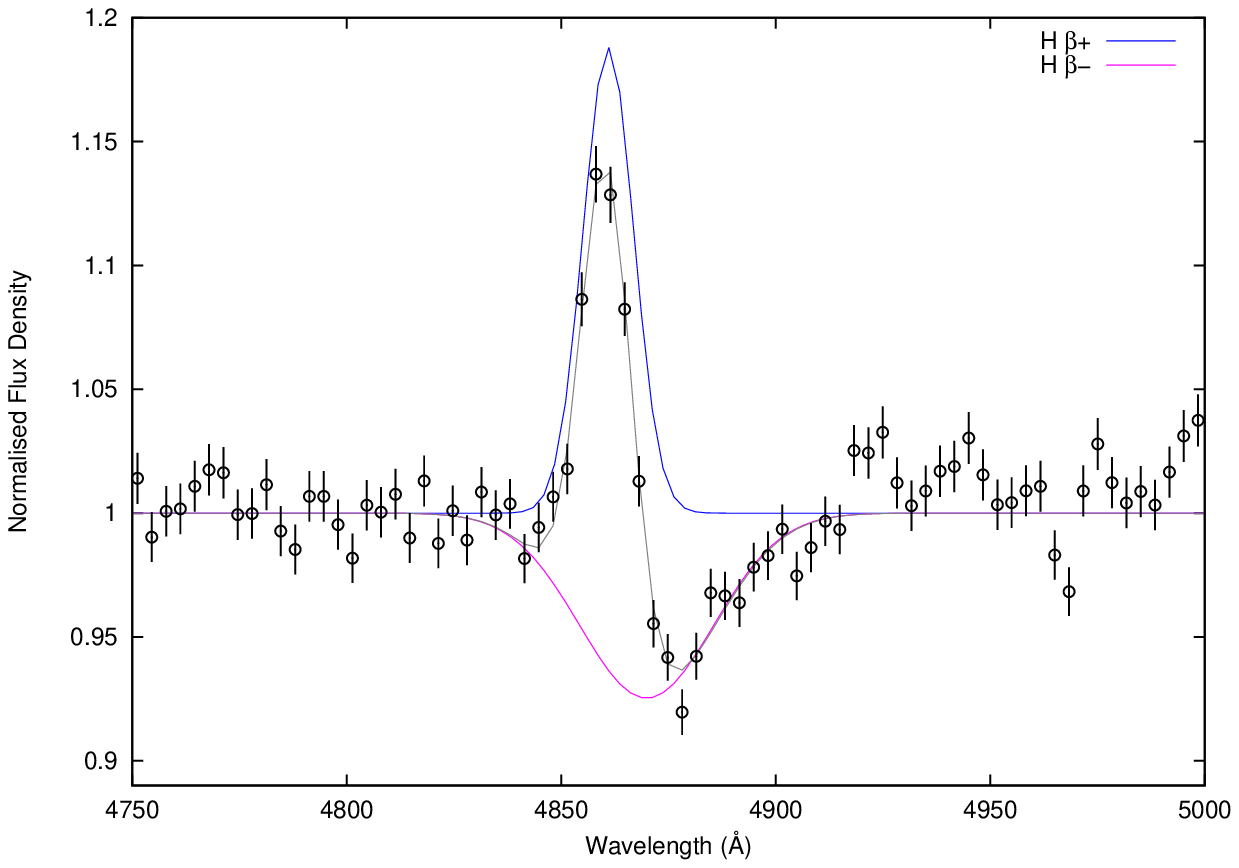}&\includegraphics[width=5.5cm]{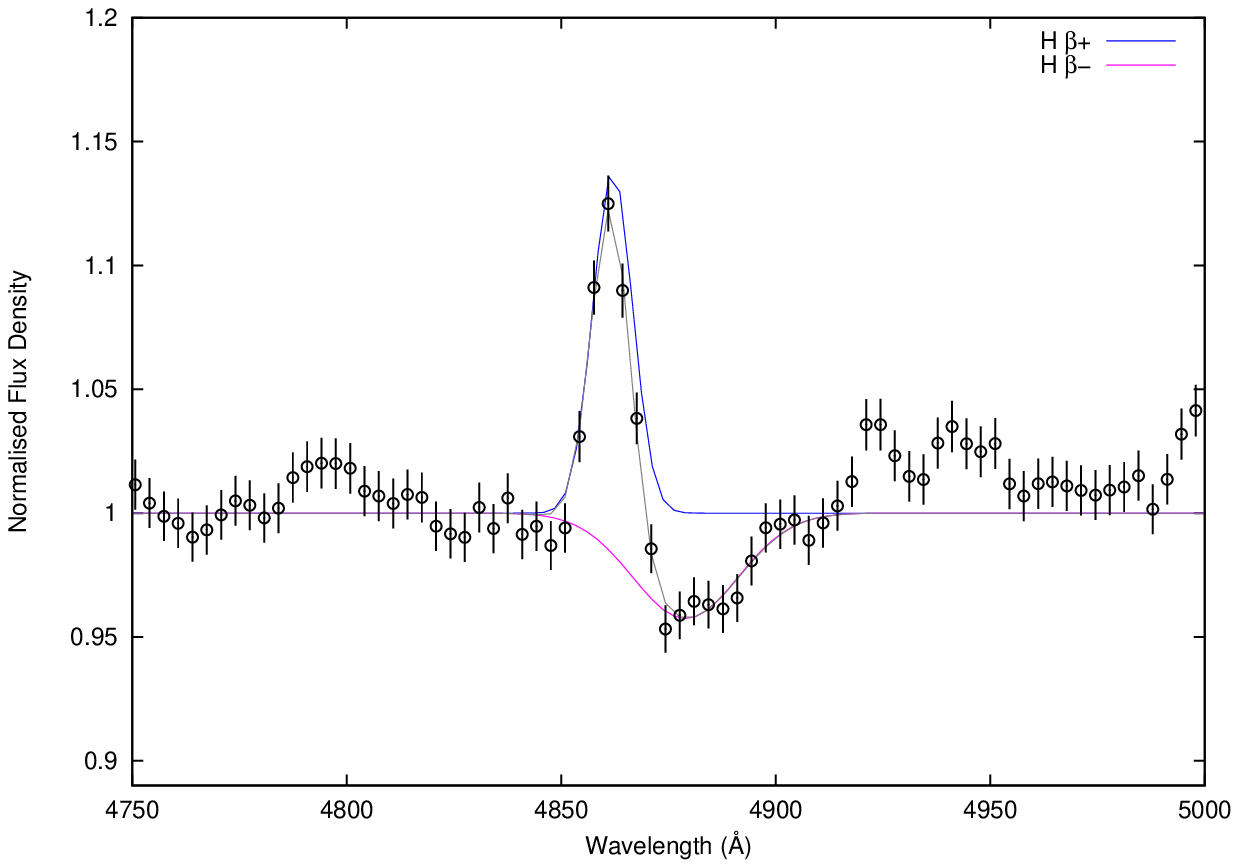}&\includegraphics[width=5.5cm]{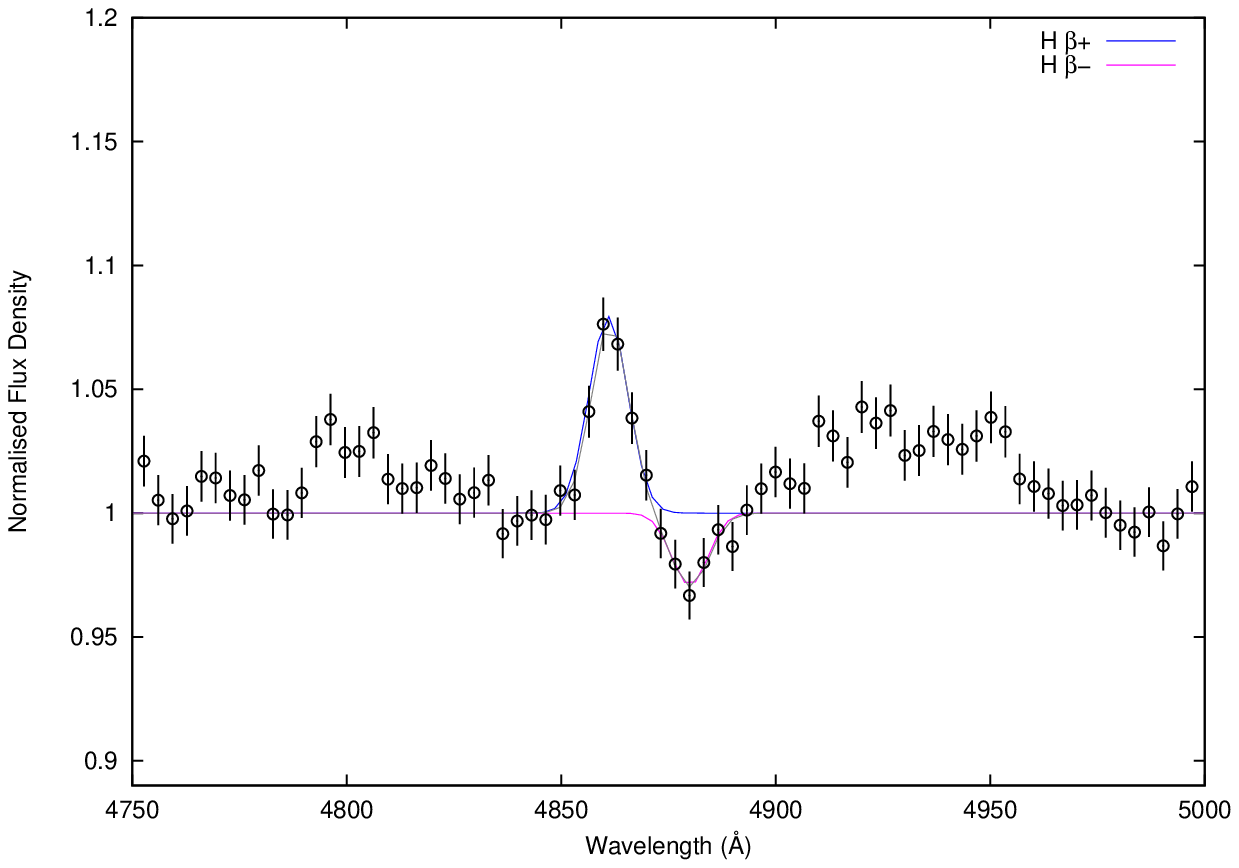}\\
\end{tabular}
\caption{\small Double-Gaussian fits of the H{\small $\beta$} feature
  for SS1 (right), HS1 (middle) and HS2 (right). During SS1 and HS1,
  the line appears to be a clear combination of a narrow emission
  (blue) and a wide absorption (magenta). During HS2, the
  detection of the trough is not significant.} 
\label{hbeta}
\end{center}
\end{figure*}

The optical spectrum is consistent with those reported in
previous studies of \gx. During SS1, we detect H{\small $\gamma$},
H{\small $\beta$}, H{\small $\alpha$} and the Paschen series, all in
emission. There are also several signatures of \ion{He}{1},
\ion{He}{2}, \ion{O}{2}, \ion{N}{2} and \ion{N}{3} (including the
Bowen Blend at 4640~\AA) although many are too weak to be properly
measured. The spectroscopic content in both HS are very similar to
that during SS1 but many of the higher ionisation features are
missing during HS2. Moreover, several unidentified lines are
detected, and two very faint emission components shortwards of
H{\small $\alpha$} appear to be present during SS1 and HS1. One is centered
at about 6474~\AA\ and has never been reported while the other,
centered at 6505~\AA, was explained as either \ion{N}{2}
\citep{1999Soria} or a violet-displaced component of H{\small
  $\alpha$} attributed to a ``jet feature'' \citep{2002Cowley}. Given
that we observe these lines in both the SS and HS, they likely
originate either in the disc or the irradiated companion
star. More importantly, we confirm the presence
during SS1, HS1 and to a lesser extent HS2 of an absorption trough
longwards of H{\small $\beta$}; its presence was already mentioned in
\citet{2003Buxton} but the authors did not conclude on its
nature. This dip significantly contaminates H{\small $\beta$} and
gives the feature an inverse P-Cygni profile. Because of this, we fit
the emission (labelled H{\small $\beta$+}) and absorption
(H{\small$\beta$-}) components simultaneously with two
   Gaussians (see Figure~\ref{hbeta}), H{\small $\beta$+} being
   centered at 4861~\AA\ and H{\small 
     $\beta$-} at 4872~\AA, 4880~\AA\ and 4881~\AA\
   during SS1, HS1 and HS2, respectively.
   All the spectral lines are also variable from one
     observation to another. Compared to SS1, their
   intrinsic fluxes are higher during HS1 and similar or slightly
   lower during HS2, and their equivalent widths are a lot lower, a
   hint that the continuum in the HS is dominated by a component that
   does not significantly enhance line emission, likely the compact
   jet. Moreover, the resolution of our spectra is too
   low to enable a sufficient analysis of the profiles of most of the
   features, but \citet{1999Smith}, \citet[][]{1999Soria} and \citet{2001Wu}
   reported several double-peaked optical lines during the soft and
   quiescent states and only one, \ion{He}{2}~$\lambda4686$, in the
   HS. In our case, one can argue that the Bowen blend during SS1,
   \ion{He}{2}~$\lambda4686$ during HS1 and the Paschen lines during
   all the observations are double-peaked. We do not see any
   significant FWHM difference in the HS compared with SS1 with
     the exception of the Bowen blend, H{\small $\beta$+} and H{\small
     $\beta$-} which are narrower and H{\small
     $\alpha$} and \ion{He}{1}~$\lambda6678$ which are broader.

\subsection{The near-IR spectrum}

The near-IR spectrum of \gx\ is very rich. During SS1, we detect
emissions from (1) Pa{\small $\beta$} in the $J$-band, (2)
the whole Brackett series (Br14 being likely blended with
\ion{He}{1}~$\lambda15859$), \ion{He}{1}~$\lambda17000$ and
\ion{He}{2}~$\lambda16243$ in the $H$-band and (3)
\ion{He}{1}~$\lambda20586, 21126$, and Br{\small $\gamma$} in the
$K$-band. Moreover, Br10, Br11, Br13 and Br14 are likely
double-peaked and this points towards the accretion disc as the main
contributor to the spectroscopic content.

In the HS, most of the aforementioned lines are present. Compared
  with SS1 their fluxes are larger during HS1 and similar during HS2,
  with the exception of  Pa{\small $\beta$} which is
  significantly brighter in both HS. In agreement with what we observe
  in the optical, their equivalent widths are however a lot lower,
  confirming that the continuum is likely dominated by the compact
  jet. In the $H$-band, Br10, Br11, Br13 and Br14 may also be
double-peaked. In the $K$-band, \ion{He}{1}~$\lambda20586$  is absent
during HS2 but present as a strong emission in a relatively deep
trough during HS1. Aside from Pa{\small $\beta$} which is broader,
there is no significant FWHM difference when compared with SS1. 

\section{Spectral analysis}\label{secan}

The wealth of \ion{H}{1} emission lines in the spectra of \gx\ allows
us to diagnose and compare the physical conditions of the  source for
each observation. For this purpose, we selected twelve of the most
robustly measured optical and near-IR \ion{H}{1} features detected in
the three spectra (bold in Table~\ref{speclines}) and discuss their
behaviour in the subsequent sections.

\subsection{The hydrogen decrements during SS1, HS1 and HS2}

We list the H{\small $\alpha$}/H{\small $\beta+$} (Balmer),
Pa($n$)/Pa11 (Paschen) and Br($n$)/Br10 (Brackett)
decrements in Table~\ref{decre}; the fluxes were corrected for the ISM
extinction using the law given in \citet{1999Fitzpatrick} with
$\Ave=3.25$. 
\begin{table}
\begin{center}
\caption{Selected Balmer, Paschen and Brackett decrements during SS1,
HS1 and HS2. Uncertainties are given at $1\sigma$.}
\label{decre}
\begin{tabular}{lccc}
\hline
\hline
&SS1&HS1&HS2\\
\hline
H$\alpha$/H$\beta+$ &0.98$\pm$0.21&1.50$\pm$0.15&2.52$\pm$0.56\\
Pa$\beta$/Pa11      &0.85$\pm$0.12&3.00$\pm$0.41&2.54$\pm$0.43\\
Pa12/Pa11     	    &0.95$\pm$0.18&0.94$\pm$0.20&0.94$\pm$0.29\\
Pa13/Pa11     	    &1.00$\pm$0.18&1.09$\pm$0.21&1.03$\pm$0.29\\
Pa14/Pa11     	    &0.98$\pm$0.20&1.12$\pm$0.22&1.07$\pm$0.38\\
Pa15/Pa11     	    &0.97$\pm$0.16&0.82$\pm$0.15&0.92$\pm$0.31\\
Br$\gamma$/Br10	    &0.59$\pm$0.10&0.59$\pm$0.10&0.63$\pm$0.14\\
Br11/Br10   	    &0.84$\pm$0.11&0.80$\pm$0.14&0.72$\pm$0.24\\	
Br14/Br10   	    &0.56$\pm$0.13&0.81$\pm$0.18&0.59$\pm$0.17\\
%\hline 
%Pa$\beta$/H$\beta+$&0.10$\pm$0.02&0.29$\pm$0.04&0.47$\pm$0.12\\
%Br$\gamma$/H$\beta+$&0.04$\pm$0.01&0.04$\pm$0.02&0.06$\pm$0.02\\
\hline
\end{tabular}

\end{center}
\end{table}

The decrements are sensitive to the physical conditions of the plasma
from which the lines arise, so that we can compare our measurements
against expected theoretical values to obtain information on the
temperature and/or density of the emitting regions. Relevant theory
comparisons are given in \citet{1980Drake} and
\citet{1983Elitzur} for high ionising fluxes while
\citet{1980Williams} and  \citet{1988Williams} provide
  information for optically thick and
thin accretion discs in cataclysmic variables (CVs). Likewise,
\citet{1995Storey} is an important source of line strength
comparisons as the authors computed the decrements for the two cases 
defined in \citet{1938Baker} for low-density nebulae that are (a)
transparent and (b) opaque to Lyman lines. Since degeneracies exist
between the plasma temperature, density and optical thickness,
we limit ourselves to a qualitative analysis. Moreover, our $J$, $H$
and $K$-band near-IR spectra are non-simultaneous. However, we showed
in Paper~I that although the continum was
variable on short timescales in the HS, it averaged out in about
100-200~s so we will assume that the same is true for the
spectroscopic features.

With that in mind, we see that the Balmer decrement 
is close to unity during SS1, significantly lower than that
for a standard Case B low density nebula
\citep[i.e. 2.86,][]{1989Osterbrock}. This is similar to what is usually
observed in CVs, an explanation being that the Balmer lines are
optically thick to the continuum and that they arise from the dense
accretion disc \citep{1980Williams}. This must be the case for \gx\
since whatever the electron temperatures and optical depth we
consider, only high densities ($N_{\rm e}\ge10^{13}\textrm{ cm}^{-3}$) 
can explain the measured small decrement values \citep[see Figures~12 and 6
in][respectively]{1980Drake, 1988Williams}. In contrast, the Balmer
decrement is higher in the HS (1.5 and 2.5 during HS1 and HS2,
respectively), which points towards lower density and optically thin
emitting regions.

The analysis of the Paschen decrements is more complex. Our
measurements are not accurate enough to get any relevant
information on the physical parameters of the emitting region. Indeed,
Pa12/Pa11 to Pa15/Pa11 are weakly dependent on temperature and
density, and a comparison with the data tabulated in
\citet{1995Storey} does not give any relevant constraints on the
physical conditions of the emitting plasma. Moreover, the Pa{\small
  $\beta$}/Pa11 values are a lot lower than the ones calculated in
\citet{1995Storey}, likely because the high ionising flux of \gx\ has
greater impact on Pa{\small $\beta$} than on the features with higher
energy levels. Nevertheless, the decrements are flat and close to unity
during SS1, which means that all the Paschen features are
optically thick to the continuum. It is important to mention that this
was already observed for Sco~X-1 \citep{1989Schachter}, likely
hinting at an accretion disc origin. During HS1 and HS2, Pa12/Pa11 to
Pa15/Pa11 are still flat and close to unity but Pa{\small
  $\beta$}/Pa11 clearly stands apart, being almost three times higher
(2.5-3). Unlike what we observe in the SS, Pa{\small $\beta$} is
therefore optically thin in the HS, which can be due
to a different emission mechanism and/or emitting region. Finally, the
Brackett decrements give similar information. The measurements are not
accurate enough for a comparison with theoretical values, but the
decrements are flat at 1$\sigma$ and point towards optically thick
lines in both the SS and HS and thus an accretion disc origin.

\begin{table}
\begin{center}
\caption{Selected Balmer, Paschen and Brackett HS1/SS1 and HS2/SS1
flux ratios. Uncertainties are given at $1\sigma$.}
\label{fr}
\begin{tabular}{lcc}
\hline
\hline
&HS1/SS1&HS2/SS1\\
\hline
H{$\beta+$}	&1.15$\pm$0.26&0.54$\pm$0.18\\
H{$\alpha$}	&1.79$\pm$0.07&1.27$\pm$0.06\\
Pa15		&1.32$\pm$0.27&1.09$\pm$0.32\\
Pa14		&1.36$\pm$0.33&1.04$\pm$0.34\\
Pa13		&1.39$\pm$0.27&0.98$\pm$0.22\\
Pa12		&1.26$\pm$0.29&0.94$\pm$0.25\\
Pa11		&1.28$\pm$0.22&0.92$\pm$0.23\\
Pa{$\beta$}	&4.30$\pm$0.42&2.75$\pm$0.27\\
Br14		&1.37$\pm$0.46&1.01$\pm$0.29\\
Br11		&1.37$\pm$0.19&0.91$\pm$0.23\\
Br10		&1.36$\pm$0.23&1.05$\pm$0.22\\
Br{$\gamma$}	&1.43$\pm$0.25&1.05$\pm$0.21\\
\hline
\end{tabular}
\end{center}
\end{table}

\subsection{Flux variations between SS1, HS1 and HS2}

Table~\ref{fr} compares the average flux ratios of the selected
\ion{H}{1} features for our three observations. The same pattern is
observed for all the Brackett and Paschen series, with the exception
of Pa{\small $\beta$}; during HS1, these lines are roughly (30-40)\%
brighter than during the SS1 and HS2 observations (which are
similar). Since the emissivity of a given line mainly
depends on its optical depth, the ionisation parameter of the emitting
plasma, the covering factor and the inclination, our finding is
consistent with all the Paschen (with the exception of Pa{\small
  $\beta$}) and Brackett features originating from the same component
during the three observations. This is in agreement with the
conclusion drawn from the decrements analysis.

The conclusion is different for the Balmer lines. Indeed, H{\small
  $\beta+$} is fainter and H{\small $\alpha$} brighter than
expected in the HS, especially during HS2. We acknowledge that the
uncertainties of our measurements prevent us from making any
definitive statement, but if true, these discrepancies are consistent
with the larger Balmer decrements measured during HS1 and HS2.
Moreover, we note that  H{\small $\beta+$} and H{\small $\alpha$}
FWHMs are respectively smaller and larger in the HS, confirming a
different and/or additional formation mechanism 
of the Balmer lines in the HS. The same conclusion can be drawn with
more certainty for Pa{\small $\beta$}, which exhibits a clear flux
excess and is a lot broader in the HS.
\begin{figure*}
\begin{center}
\begin{tabular}{cc}
\includegraphics[width=8cm]{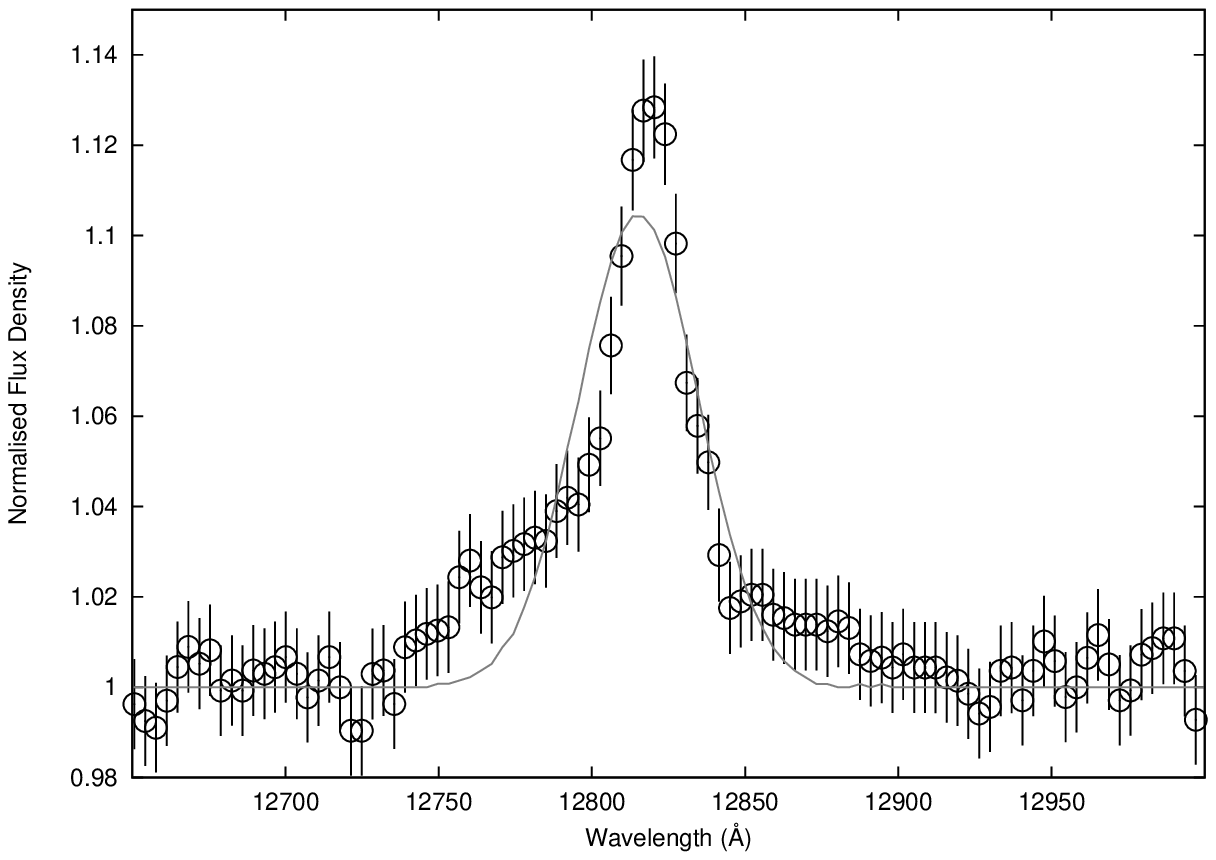}&\includegraphics[width=8cm]{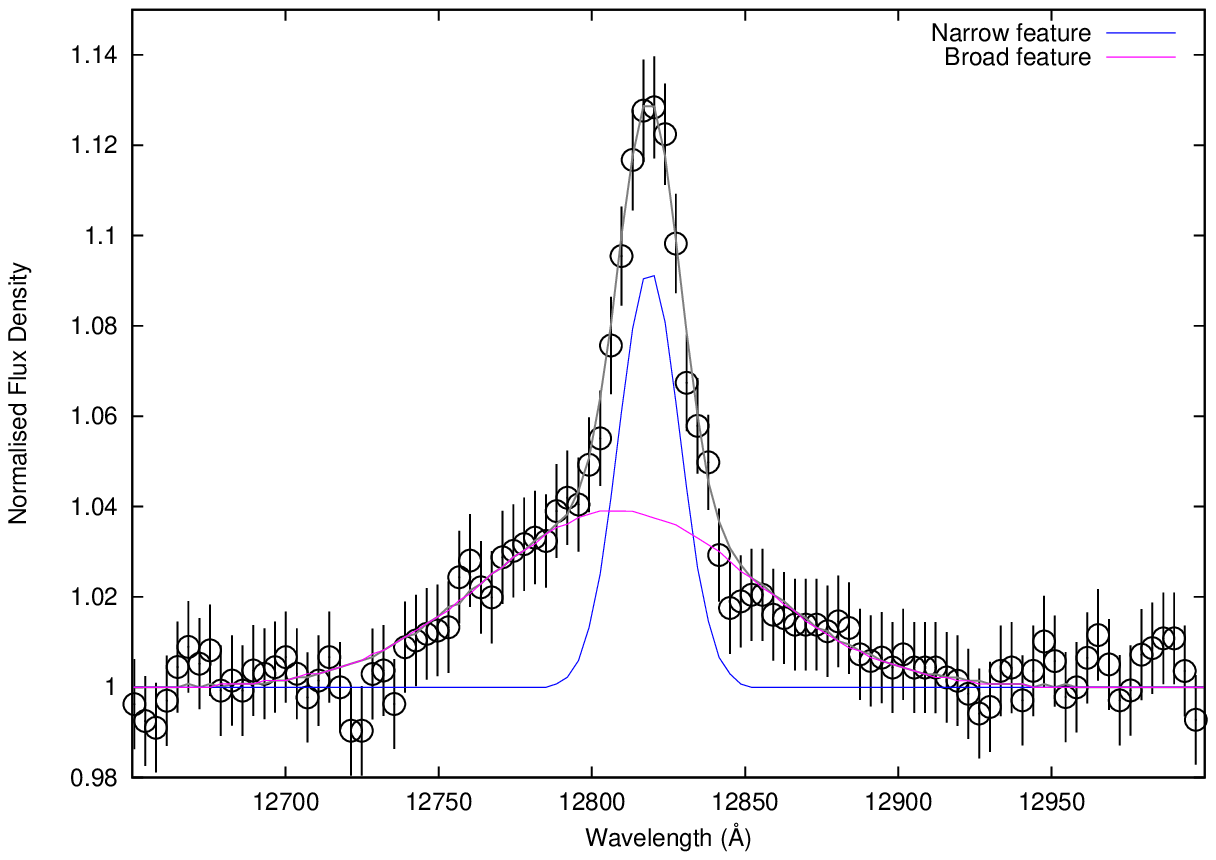}\\
\includegraphics[width=8cm]{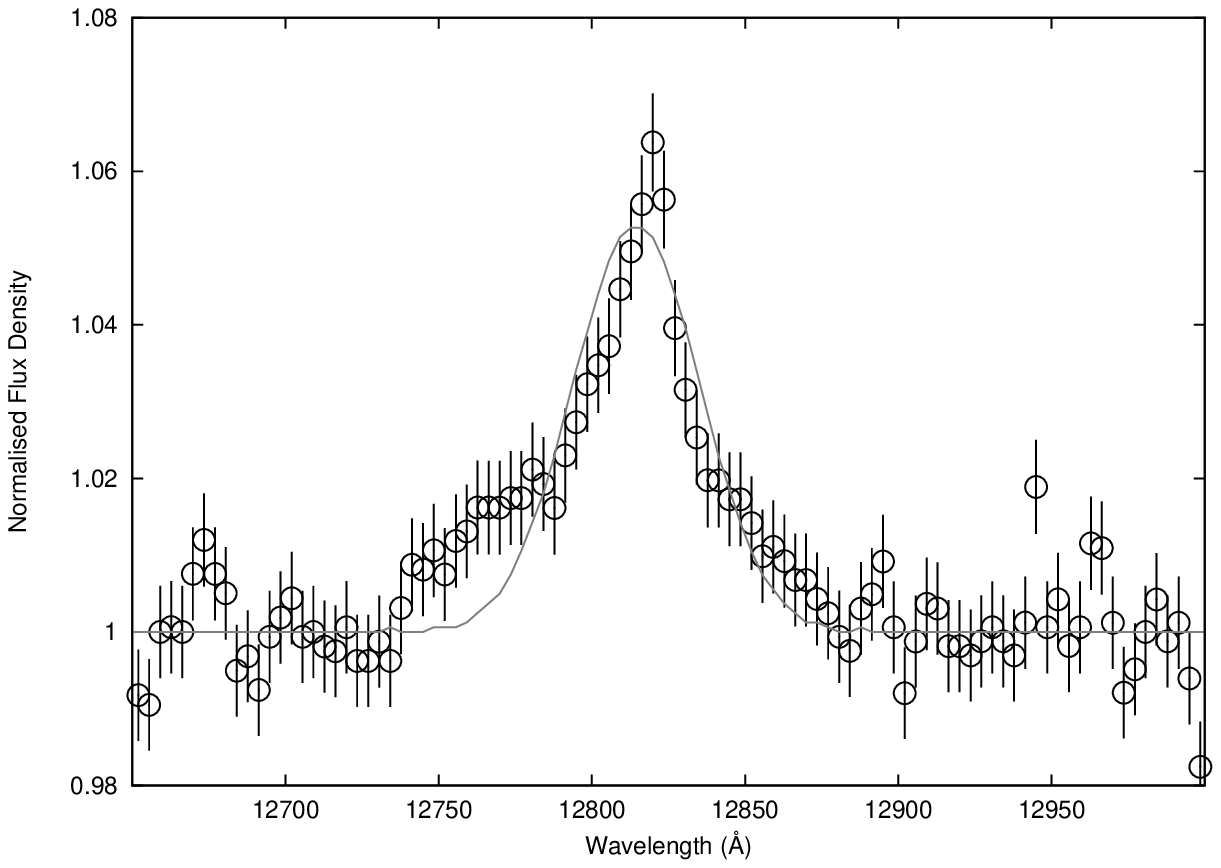}&\includegraphics[width=8cm]{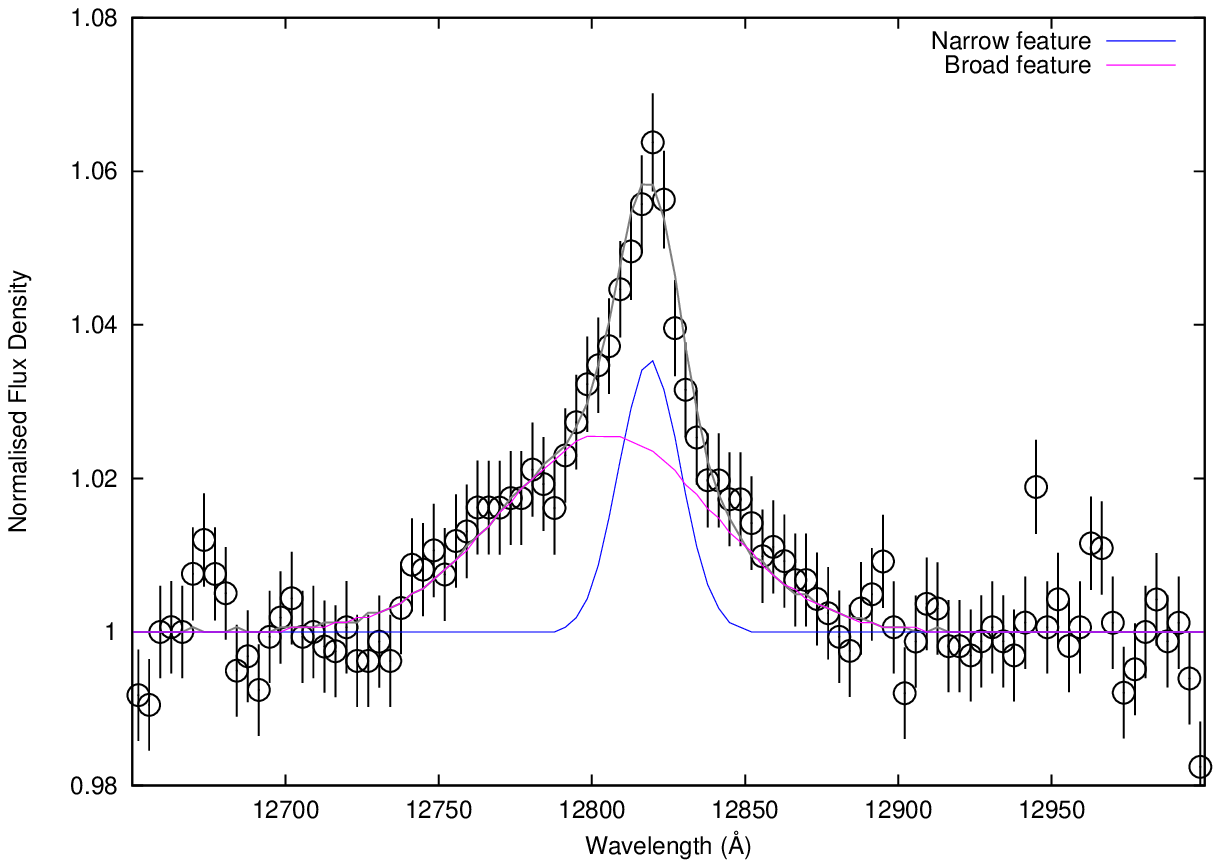}\\
\end{tabular}
\caption{\small Pa{\small $\beta$} fitted with one (left), 
  and two (right) Gaussian(s) during HS1 (top) and HS2
  (bottom). Uncertainties are give at $1\sigma$.}
\label{pbetaprof}
\end{center}
\end{figure*}
\section{Discussion}\label{secdis}

In the OIR domain, double-peaked emission lines thought to arise from a
temperature-inversion layer created by the soft X-ray irradiation of
the chromosphere \citep[see e.g.][]{1980Jameson,1986Horne,2001Wu} have
long been associated with microquasar accretion discs. Based on
our observed flat decrements, Keplerian FWHMs, homogeneous flux
variation patterns and, to a lesser extent, double-peaked profiles, an
irradiated accretion disc is the likely origin for all the
\ion{H}{1} lines we observe. Those for which this is not the case are discussed
subsequently.

\subsection{Origin of the H{\small $\beta$} absorption trough}

Of primary interest is the presence of H{\small $\beta$-}. A
straightforward explanation is that it is associated with the
4882~\AA\ diffuse interstellar band \citep[DIB, see][for a
review]{1995Herbig} and this may indeed be the case during HS2, as the
trough is barely detected. DIBs however are expected to be narrow and
to have a constant equivalent width\footnote{As for the DIBs we
  unambiguously detect, like the doublet centered at 5780~\AA\
  ($1.08\pm0.13$~\AA, $0.99\pm0.18$~\AA\ and $1.01\pm0.10$~\AA\ for
  SS1, HS1 and HS2, respectively), NaD at 5890~\AA\ ($3.94\pm0.09$~\AA,
  $3.93\pm0.17$~\AA\ and $3.91\pm0.16$~\AA) or the DIB located at
  6284~\AA\ ($1.72\pm0.12$~\AA, $1.75\pm0.18$~\AA\ and
  $1.77\pm0.16$~\AA).}, yet the FWHMs of H{\small $\beta$-}
during SS1 and HS1 are clearly consistent with a strong
rotational broadening and its equivalent width varies between these
two observations. We therefore argue that during SS1 and HS1, H{\small
  $\beta$-} is mainly an absorption signature of H{\small $\beta$}
that stems from a region closer to the black hole, with a weak
contribution of the 4882~\AA\ DIB. In constrast, the weak absorption
detected during HS2 is not associated with \gx\ but only with the
4882~\AA\ DIB; in other words, the H{\small $\beta$} absorption line
detected during SS1 and HS1 is absent during HS2. Balmer line
absorption troughs can form in optically thick accretion discs with a
vertical temperature decreasing with height above the central plane
\citep{1989Ladous}. We note that such troughs are common in dwarf
novae and this is the reason why we favour an accretion disc origin
for H{\small $\beta$-}. This hypothesis was also preferred to explain the
presence of such broad H{\small $\beta$} absorptions in other
outbursting microquasars like GRO~J0422+32 \citep{1995Callanan},
GRO~J1655-50 \citep{1997Bianchini, 2000Soria} or XTE~J1118+480 
\citep{2001Dubus}. For our case, H{\small $\beta$-} may
be slightly red-shifted and we refer the interested reader to
\citet{1995Callanan} and \citet{2001Dubus} for a discussion of
possible explanations for the red-shift, which include relativistic
effects or superhump-induced distortions.

\subsection{An irradiation-induced envelope in the HS}

A comprehensive modelling of the conditions for Balmer absorption
  line formation in black hole X-ray binaries was presented in
\citet{1998Sakhibullin}. As stated in the previous section, H{\small
$\beta$-} is strong during SS1 and HS1 but absent during HS2. Of
relevance to our findings, the Balmer absorption lines are more
prominent when at least one of the following conditions is fulfilled:
(1) the X-ray emission is harder, (2) the degree of self-illumination
(i.e. fraction of reprocessed X-rays) is lower, or (3) the accretion
rate is lower. (1) The 3--50~keV \rxte\ spectrum of \gx\
 is however as hard during HS1 as HS2, with a 6--50~keV to 3--6~keV
 PCA hardness ratio of about 1.45 and 1.47 during HS1 and HS2,
 respectively (see Paper~I);  we can therefore exclude the hardness of
 the X-ray emission as responsible  for the disappearance of H{\small
   $\beta$-} during HS2. (2) The same conclusion can be drawn for X-ray
 irradiation, the lower optical continuum during HS2 hinting towards a
 lower level of X ray reprocessing, which should be more favourable to
 absorption line formation. (3) Finally, it is very unlikely that the accretion
rate during HS2 (1.2-3~keV ASM/flux of about
2~counts~cm$^{-2}$~s$^{-1}$) is higher than during SS1 (1.2-3~keV
ASM/flux of about 27~counts~cm$^{-2}$~s$^{-1}$). The disappearance of 
H{\small $\beta$-} during HS2 must therefore stem from another
mechanism and the Pa{\small $\beta$} properties in the
  HS are key to understanding this behaviour.
\begin{table*}
\begin{center}
\caption{Derived parameters from two Gaussian (narrow+broad) fit to Pa{\small
$\beta$} during HS1 and HS2. Uncertainties are given at 1$\sigma$.}
\label{multgauss}
\begin{tabular}{cccc|cccc}
\hline\hline
\multicolumn{4}{c|}{HS1}&\multicolumn{4}{c}{HS2}\\
\hline
{$\lambda_{\rm \small c}$}$^a$&$FWHM^b$&$\dot{W}^c$&$Flux^d$&{$\lambda_{\rm \small c}$}&$FWHM$&$\dot{W}$&$Flux$\\
\hline
12815&2292$\pm$510&3.32$\pm$0.91&8.72$\pm$1.25&12809&1735$\pm$486&2.00$\pm$0.56&6.00$\pm$1.13\\
12818&515$\pm$231&1.74$\pm$0.91&4.40$\pm$0.80&12819&502$\pm$256&0.89$\pm$0.41&2.51$\pm$0.54\\
\hline
\end{tabular}
\end{center}
{{$^a$}{Derived wavelength {(\AA)}} \\}
{{$^b$}{Full-width at half-maximum} (\kms)\\}
{{$^c$}{Equivalent width} (\AA)\\}
{{$^d$}{{Line flux} ($\times 10^{-15}$erg cm$^{-2}$ s$^{-1}$)}\\}
\end{table*}

\begin{figure*}
\begin{center}
\begin{tabular}{cc}
\includegraphics[width=8cm]{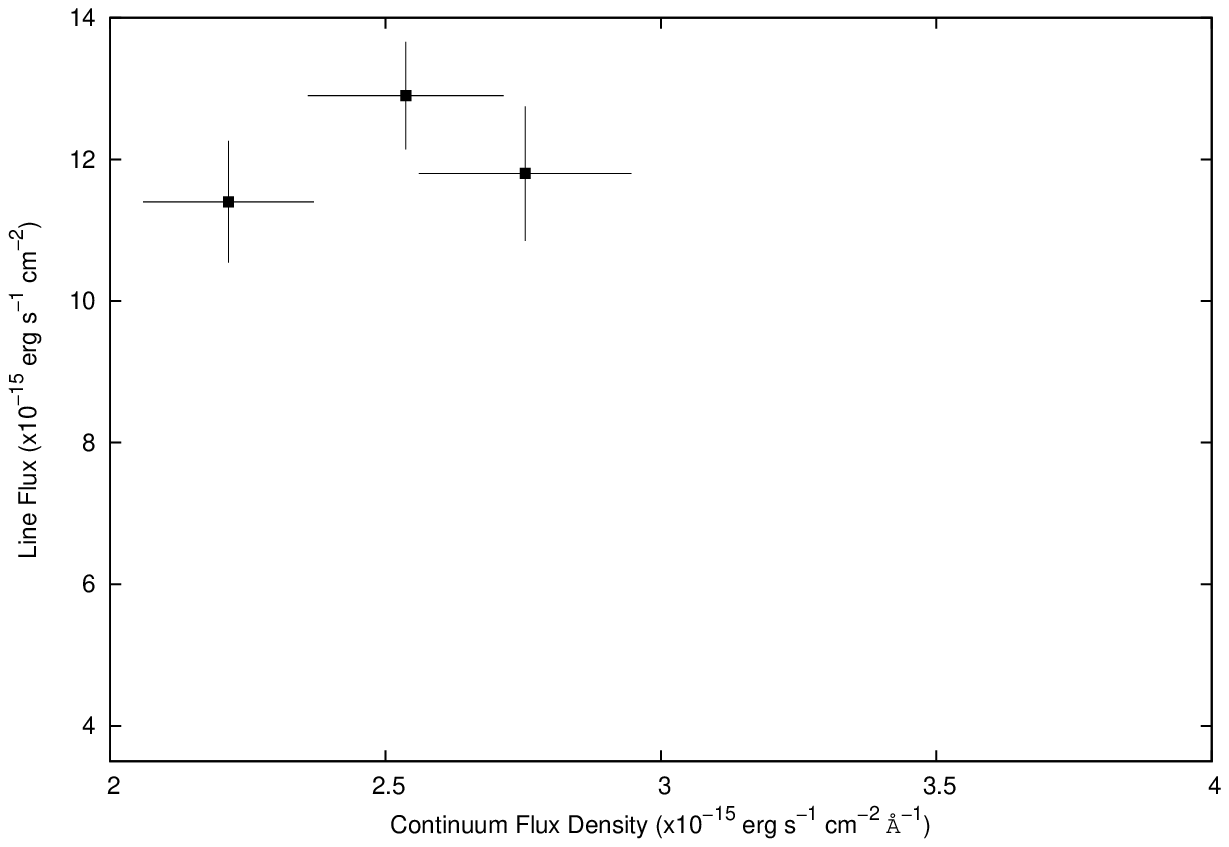}&\includegraphics[width=8cm]{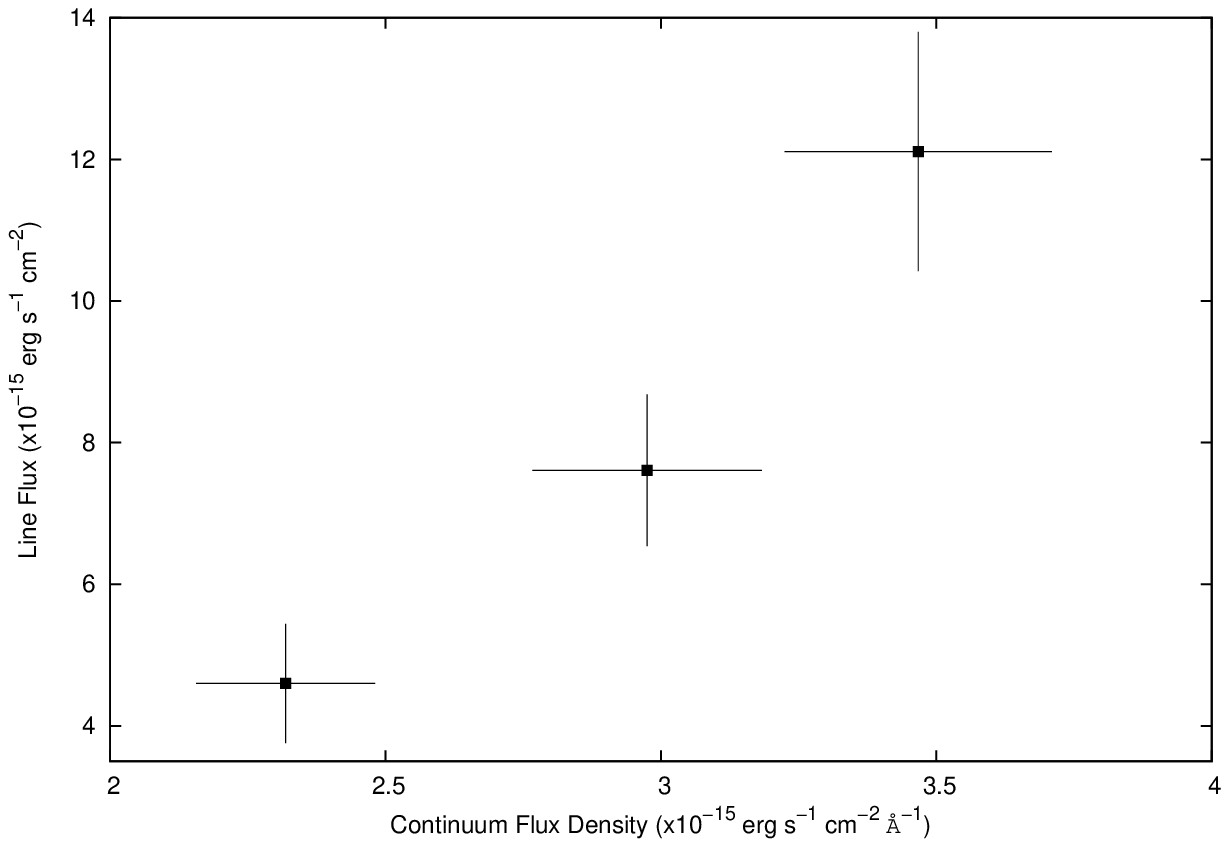}\\
\end{tabular}
\caption{\small Pa{\small $\beta$} flux variations as a function of the
  local continuum flux density during HS1 (left) and HS2
  (right). Uncertainties are given at 1$\sigma$.}
\label{fluxvar}
\end{center}
\end{figure*}

Indeed, compared to the other near-IR \ion{H}{1} emission features,
Pa{\small $\beta$} exhibits a strong flux excess and is twice broader
in the HS compared to the SS. It is is thus reasonable to assume that
several components contribute to the emission. To test this
hypothesis, we fitted the HS1 and HS2 Pa{\small $\beta$} profiles
with two Gaussians (see Figure~\ref{pbetaprof} for a comparison with
the one Gaussian case). We performed Fisher's tests to assess the need
of a second Gaussian (degree of freedom of 350 for the two-Gaussian
case vs 353 for the one-Gaussian case)  and obtained $10^{-22}$ and
$10^{-6}$ chance probabilities for HS1 and HS2, respectively. The
  need for two Gaussians is therefore statistically justified and
  Pa{\small $\beta$} is better described by the combination of a
  narrow feature likely coming from the outer accretion disc and a
  broad one, hereafter Pa{\small $\beta_{\rm b}$}, arising from
  another region (their derived parameters for both HS1 and HS2 are
  listed in Table~\ref{multgauss}). Assuming that the broadening of
  Pa{\small $\beta_{\rm b}$} is mainly due to rotation, the typical
  distance to the black hole $R_\beta$ of the region that contributes
  the most to the emission can be inferred from its FWHM. Indeed, we have
  $R_\beta=\frac{1}{2}\left(\frac{c}{V_\beta}\right)^2 R_{\rm   S}$,
  where $V_\beta$ is the Keplerian velocity and $R_{\rm S}$ 
the Schwarzschild radius. We find $V_\beta {\rm sin}\,i\sim1375$~\kms\ 
and $V_\beta {\rm sin}\,i\sim1041$~\kms\ as well as
$R_\beta\sim2.4\times10^4 ({\rm sin}\,i)^2 R_{\rm S}$ and
$R_\beta\sim4.2\times10^4 ({\rm sin}\,i)^2 R_{\rm S}$ during HS1 and
HS2, respectively ($i$ being the system inclination). Using the
orbital parameters derived in \citet{2003Hynes}, i.e. the mass
function $f(M_{\rm X})\sim5.8$~\msun, the companion star to black hole
mass ratio $q\sim0.08$ and the orbital period $P\sim1.76$~days, we
estimate $a\sim4.6\times10^5({\rm sin}\,i)^2 R_{\rm S}$, $R_{\rm
  circ}\sim0.34a$ and $R_{\rm tide}\sim0.54a$, where $a$ is the
orbital radius,  $R_{\rm circ}$  is the circularisation radius and
$R_{\rm tide}$ is the tidal radius \citep[see][for the expression in function
of $q$]{2002Frank}. Because the outer radius of the accretion disc
$R_{\rm out}$ is larger than $R_{\rm circ}$ and smaller than $R_{\rm
  tide}$, we have $1.6\times10^5({\rm sin}\,i)^2 R_{\rm S}\le R_{\rm
  out} \le 2.5\times10^5({\rm sin}\,i)^2 R_{\rm S}$. The inclination
of the system is unclear, but the most recent measurement points
towards $i\sim50^{\circ}$ \citep{2011Shidatsu}, which results in $9.1\times10^4R_{\rm
  S} \le R_{\rm out} \le 1.5\times10^5 R_{\rm S}$ as well as $R_\beta =
(1.4\pm0.6)\times10^4R_{\rm S}$ and $R_\beta=
(2.5\pm1.4)\times10^4R_{\rm S}$ during HS1 and HS2, respectively.

The fact that we still detect narrow emission lines in the HS1 and HS2
near-IR spectra of \gx\ means that the optically thick and
geometrically thin outer disc is still present, whereas Pa{\small
  $\beta_{\rm b}$} arises from an extended region that covers a
significant fraction of the inner accretion disc. While such a
configuration is similar to truncated discs/hot 
inner flow geometries thought to be ubiquitous in the HS
  \citep[e.g. advection-dominated accretion flows or
  ADAF,][]{1995Narayan}, $R_\beta$ is too large to be  consistent with 
  the transition radii between the optically thin and thick regions
  derived in these models. For instance, such high radii are predicted
  for ADAFs with very low accretion rates
  $\dot{M}\le10^{-4}\dot{M}_{\rm edd}$, typical of quiescent  
  states. Likewise, we were able to model the \sw/XRT+\rxte/PCA
  spectrum of \gx\ during HS1 with an accretion disc model extending
  close to the black hole and we found $R_{\rm in}\sim3.9-13.2R_{\rm
    S}$ for $i=50^{\circ}$, $d=8$~kpc and a 10~\msun\ black hole
  mass. We thus rather propose that in the HS only, the hard X-ray
  emission irradiates the chromosphere of the accretion disc which
  eventually puffs up into a geometrically thick envelope that can
  enshroud the region of the accretion disc from which H{\small
    $\beta$-} originates. We estimate that the typical distance to the
  black hole of the region that contributes the most to H{\small
    $\beta$-} during HS1 is about $(2.6\pm1.2)\times10^4R_{\rm S}$, to
  be compared to a typical radial extension of the envelope of
  $(1.4\pm0.6)\times10^4R_{\rm  S}$ and $(2.5\pm1.4)\times10^4R_{\rm
    S}$ during HS1 and HS2, respectively. Two scenarios are thus
  possible to explain the H{\small $\beta$-} detection during HS1 and
  its disappearance during HS2, but the accuracy of our measurements
  prevents us to disentangle between them:
\newline
(1) the envelope is optically thick during the two HS. H{\small
  $\beta$-} is therefore undetected during HS2 because the envelope
inflates and enshrouds the region that contributes the most to the
feature,  
\newline
(2) the envelope enshrouds the H{\small $\beta$-} region during the
two HS.  H{\small $\beta$-} is then detected during HS1 because
the envelope is optically thin, while it becomes optically thick
during HS2.
\newline

Finally, it is likely that a significant fraction of H{\small 
  $\alpha$} comes from this envelope, which would then be
optically thick to the Balmer series \citep[Case C
recombination,][]{1992Xu}. Indeed, each H{\small $\beta$} and higher 
Balmer lines would in this case be split into H{\small $\alpha$} plus
some near-IR recombination features such as H{\small
  $\beta$}$\rightarrow$H{\small $\alpha$}+Pa{\small $\alpha$},
H{\small $\gamma$}$\rightarrow$H{\small $\alpha$}+Pa{\small $\beta$}, 
etc. \citep[See][for a similar statement in the case of
GRO~J1655-40]{2000Soria}. This would be in agreement with the brighter
H{\small $\alpha$}, fainter H{\small $\beta$+} and larger Balmer
decrements encountered in our HS observations, especially during HS2.

\subsection{A thermal accretion disc wind in the HS?}

We showed in Paper~I that the near-IR continuum was
variable at 20~s timescales during HS1 and HS2 but constant during
SS1. It is therefore relevant to investigate the Pa{\small $\beta$}
variability with respect to that of the continuum, but S/Ns are 
too low for us to evaluate it on 20~s
timescales. Instead, we combined the individual $J$-band spectra in
groups of four by decreasing continuum level, measured the Pa{\small
  $\beta$} flux in each of the three resultant 80~s spectra and
plotted them as a function of the underlying continuum level (see
Figure~\ref{fluxvar}). It is clear that Pa{\small $\beta$} is constant
during HS1 and correlated with the continuum during HS2. We followed
the same procedure for Pa{\small $\beta$} during SS1 and
Br{\small $\gamma$} during the three observations and find the intrinsic
fluxes to be constant at 1$\sigma$, a hint that the emission lines
coming from the accretion disc are constant. Thus, to understand if
Pa{\small $\beta_{\rm b}$} was indeed responsible for the flux
variations during HS2, we combined the twelve 20~s $J$-band spectra 
into two 120~s sub-spectra $-$ either with a high or a low continuum
level $-$ and we measured the respective contributions of the narrow
and broad Pa{\small $\beta$} components in each of them. For the high
continuum level spectrum, we find that Pa{\small
  $\beta$} can be well-described by a
$4.85\pm2.55\times10^{-15}$~\ergcms\ narrow feature and a
$6.72\pm3.34\times10^{-15}$~\ergcms\ broad one. In constrast, the same
fit for the low continuum level spectrum gives
$3.90\pm2.08\times10^{-15}$~\ergcms\ for the narrow feature and
$2.80\pm1.50\times10^{-15}$~\ergcms\ for the broad one. Although not
conclusive, these measurements point towards the broad component 
being responsible for the Pa{\small $\beta$} variations. If this is indeed
the case, Pa{\small $\beta_{\rm b}$} must be optically thick to the
variable continuum during HS2, which either implies (1) a change of 
the intrinsic properties of the envelope which becomes optically
thick to the compact jet emission, the latter being responsible for the
continuum variations or (2) the near-IR continuum during HS2 is
dominated by the envelope and not the compact jets. Unfortunately,
we cannot disentangle between these scenarios, but both explanations
could account for the peculiar shape of the \gx\ spectrum during HS2
(see Figure~\ref{contspec} in the present paper and Figure~5 in
Paper~I) and might be consistent with the presence of a
  thermally-driven accretion disc wind (ADW) during HS2.

Indeed, the presence of a geometrically thick envelope in the HS is
consistent with what was proposed in 
\citet{1983Begelman} to describe the conditions necessary to form
thermally-driven ADWs. Moreover, the presence of ADWs in the HS of
\gx\ has already been proposed to explain the flat-topped or
round-topped profiles of the Balmer emission lines detected through
high-resolution optical spectroscopy \citep{2001Wu}. So, can the \gx\
envelope really launch such thermal ADWs in the HS? According to
\citet{1983Begelman}, this is possible for a launching radius $R_{\rm
  L}\ge0.2\times R_{\rm C}$ where 
\begin{equation}
R_{\rm C} \,=\,2.5\times10^{-15}\frac{c^2}{GT_{\rm C}}R_{\rm S}
\end{equation}
is the Compton radius and 
\begin{equation}
T_{\rm C}\,=\,\frac{1}{4k_{\rm B}L}\int_{\nu_1}^{\nu_{\rm N}} h\nu L_{\nu}
  d\nu\,\,\textrm{K} 
\end{equation}
is the Compton temperature \citep[see also][for a discussion in the case
of GRS~1915+105]{2010Rahoui}. This thermal ADW must moreover overcome
gravity, which can be achieved if the luminosity $L$ of the
  irradiating source is about twice
larger than the critical luminosity $L_{\rm CR}$ defined as 
\begin{equation}
L_{\rm CR}\,=\,\frac{288}{\sqrt{T_{\rm C}}}L_{\rm
  edd}
\end{equation}

Based on the modelling of the \sw/XRT+\rxte/PCA spectra of \gx\
during HS1, we estimate $T_{\rm  C}\approx6.6\times10^7$~K, $R_{\rm
  C}\approx5.1\times10^4R_{\rm S}$, $L\approx0.05L_{\rm edd}$ and
$2\times L_{\rm  CR}\approx0.07L_{\rm edd}$. We stress that these
calculations are model-dependent beyond 100~keV, where the slope
differs depending on whether the hard X-ray emission is modelled with
a power law, a cut-off power law or a Comptonisation component; this
is the reason why we integrated the spectrum between 0.1 and 100~keV
only. Furthermore, the lack of \sw/XRT data prevents us for better
modelling the 0.1-100~keV emission of \gx\ during HS2, but we consider
that the $R_{\rm C}$, $T_{\rm C}$ and $L_{\rm CR}$
measurements for HS1 are also valid for HS2 due to the similarity of
the hard X-ray emission in both HS. That said, the conditions could be
met for a thermal ADW to arise during HS1 and HS2 if the geometrically
thick envelope is larger than $R_{\rm 
  L}\ge1\times10^4R_{\rm S}$. We previously estimated that the size of
the envelope is about $(1.4\pm0.6)\times10^4R_{\rm S}$ during HS1
and $(2.4\pm1.3)\times10^4R_{\rm S}$ during HS2. Although possible in
both HS within uncertainties, it is thus clear that the conditions are
more favourable for a thermal ADW to be launched during HS2.
\newline

Interestingly, the conditions for a thermal ADW in the SS of \gx\ are
not met. Indeed, we do not detect any spectral signature for an
geometrically thick envelope in the OIR
spectrum during SS1, which means that it is likely absent in the
SS. Yet, the calculations for the conditions needed for a thermal ADW
during SS1 result in $R_{\rm L}\ge6.6\times10^4R_{\rm S}$, i.e. a
launching radius about seven times larger than in the HS. Thus, if
these calculations are correct, it is impossible for a thermal ADW to
arise during SS1. As such, another mechanism must  be responsable for
the formation of the ADW in the SS of \gx, plausibly magnetic driving
\citep[see][for the case of GRO~J1655$-$40]{2008Miller}.

\section{Conclusion}\label{seccon}

The OIR spectra of \gx\ in the two HS have different properties, and
we showed that during HS2, (1) a near-IR continuum excess is present,
(2) the H{\small $\beta$} absorption component is absent, (3) the
H{\small $\alpha$}/H{\small $\beta$} decrement is larger than that during
HS1 and (4) Pa{\small $\beta$} is variable and correlated with the
underlying continuum. Although speculative, it is tempting to
associate these differences with the presence of a thermal ADW
launched by the geometrically thick envelope during
HS2. Without definitive detections of ``typical wind signatures'' such
as blue-shifted emission lines or P-cygni profiles -- which would
require higher spectral resolutions -- we cannot ascertain the
aforementioned statement. We can nonetheless speculate on the
implications this scenario could have for our understanding of
accretion/ejection processes in \gx. From a soft X-ray prospective, we
know that an anti-correlation is observed between
compact jets in the HS and blue-shifted absorption lines of ionised
H-like species in the SS, considered as disc wind signatures. This led
some authors to speculate that both outflow types compete for the same
matter supply \citep[see e.g.][]{2009Neilsen, 2012Ponti}. 
However, the possible presence of a thermal ADW in the HS and a
magnetic ADW in the SS may rather imply a three-player game and
complicate the situation \citep[see][for the case of
GRO~J1655$-$40]{2012Neilsen}. For instance, if the compact jets are
highly-collimated and accelerated magnetic ADWs \citep{1982Blandford,
  1983Pudritz}, the question then is whether a thermal ADW arising in
the HS because of hard X-ray irradiation can significantly impact the
configuration of a large scale magnetic field to decrease the degree
of collimation through angular momentum losses. These considerations
are obviously premature and a first step before proceeding further is
to confirm the presence of thermal ADWs in the HS of microquasars. In
that respect, optical and near-IR high-resolution spectroscopy could
be a good alternative to the soft X-ray domain and we recommend
further OIR HS observations of outbursting microquasars with
instruments such as ESO/X-shooter or Magellan/FIRE to further explore
this point.

\section*{Acknowledgments}
We thank the anonymous referee for his/her very constructive review,
which significantly improved this paper. FR thanks the ESO staff who
performed the service observations. JCL thanks the Harvard Faculty of
Arts and Sciences and the Harvard College Observatory. This research
has made use of data obtained from the High Energy Astrophysics
Science Archive Research Center (HEASARC), provided by NASA's Goddard
Space Flight Center. This research has made use of NASA's Astrophysics
Data System, of the SIMBAD, and VizieR databases operated at CDS,
Strasbourg, France.

\bibliographystyle{mn2e}
\bibliography{./mybib}{}

\label{lastpage}

\end{document}